# GENAI DETECTION TOOLS, ADVERSARIAL TECHNIQUES AND IMPLICATIONS FOR INCLUSIVITY IN HIGHER EDUCATION

## A PREPRINT

Mike Perkins [1*], Jasper Roe [2], Binh H. Vu [1], Darius Postma [1], Don Hickerson [1], James McGaughran[1], Huy Q. Khuat [1]

[1] British University Vietnam, Vietnam.
[2] James Cook University Singapore, Singapore.
*Corresponding Author: Mike.p@buv.edu.vn

March, 2024

## Abstract

This study investigates the efficacy of six major Generative AI (GenAI) text detectors when confronted with machine-generated content that has been modified using techniques designed to evade detection by these tools (n=805). The results demonstrate that the detectors' already low accuracy rates (39.5%) show major reductions in accuracy (17.4%) when faced with manipulated content, with some techniques proving more effective than others in evading detection.

The accuracy limitations and the potential for false accusations demonstrate that these tools cannot currently be recommended for determining whether violations of academic integrity have occurred, underscoring the challenges educators face in maintaining inclusive and fair assessment practices. However, they may have a role in supporting student learning and maintaining academic integrity when used in a non-punitive manner.

These results underscore the need for a combined approach to addressing the challenges posed by GenAI in academia to promote the responsible and equitable use of these emerging technologies. The study concludes that the current limitations of AI text detectors require a critical approach for any possible implementation in HE and highlight possible alternatives to AI assessment strategies.

## Introduction

Recent developments in Artificial Intelligence (AI) have led to paradigm-shifting applications and technologies which have significantly impacted multiple aspects of society. The 'AI Spring' (Manyika & Bughin, 2019) of the early 2020s and its ongoing consequences has already garnered a great deal of attention in academia and education in a short span of time. At the time of writing, AI is receiving greater adoption both formally and informally among students and teachers and has become a 'hot topic' (Cotton et al., 2023), of which our understanding is only just beginning. Generative AI (GenAI) tools, that is, those which produce an output of some form in response to user input, whether text, audio, image, or video, have attracted the most attention in this area. Despite many proclaiming the potential benefits of these tools in aiding learning, GenAI also presents several risks, including equity and inclusion in education. Despite such risks, a consensus on the acceptance of GenAI in some spheres of academia is emerging. For example, many academic publishers have moved towards a position which accepts the use of GenAI in crafting manuscripts, albeit with certain restrictions (Perkins & Roe, 2024a), and studies have shown the rising popularity of using GenAI tools to support academic writing and research (Bedington et al., 2024; Chan, 2023; Perkins & Roe, 2024b; Sobaih, 2024). Simultaneously, a growing number of organizational bodies are now producing guidelines for the ethical use of GenAI in education (Foltynek et al., 2023; Miao & Holmes, 2023).

Regarding inequality and inclusion, the consensus acceptance of GenAI tools in scientific publishing may disadvantage certain groups of students and researchers. For example, barriers in access to the internet, financial barriers in accessing paid GenAI tools (as premium versions often require a subscription) and other issues of access (for example, disability) all play a role in worsening 'digital poverty' (Miao & Holmes, 2023). Further, Bissessar (2023) also identifies the digital divide as an important consideration in the use of GenAI tools in classroom settings and Liang (2023) recognizing that GenAI tools are known for producing structurally biased and ethnocentric worldviews.

This is not the only concern regarding GenAI technology in HE. Shortly after the rapid increase in the popularity of GenAI tools caused by the release of ChatGPT in November 2023, tools were released commercially, claiming that textual GenAI outputs could be detected. We refer to these as GenAI text detectors. Many of these detectors claim a high degree of reliability in identifying the presence of AI-generated content in submissions, often giving either a percentage score (i.e. 'this text is 10% AI-generated') or a likelihood score (i.e. 'this text is likely to be AI-generated'). Underlying the principle of these detectors is the assumption that by identifying GenAI text, assessors and educators can ensure fair and secure assessment. Research has shown that the assuredness of fairness and equity is vital for inclusive assessment practices (Morris et al., 2019), and notions of fairness and security suggest that students will not be able to misrepresent authorship and thus gain educational outcomes that they have not earned.

However, research has shown that GenAI text detectors have the potential to be barriers to inclusive assessment practices by disproportionately targeting individuals who don't speak English as their first language, or those with lower English proficiency (Liang et al., 2023). In this study, we use the term 'Non-Native English Speaker' (NNES) to refer to this population. We recognize that there are various terms used to describe individuals who don't speak English as their first language, such as English Language Learner (ELL), English as a Second Language (ESL) speaker, and English as an Additional Language (EAL) speaker with each term having its own merits and limitations. While NNES may be seen as problematic by some because it categorises people based on what they are not, rather than what they are, we have chosen to use this term in our study because our primary focus is on exploring biases specifically related to these groups, and in this context believe NNES to be the most accurate and direct term to describe our target population.

It has long been known that non-native English speakers face challenges when participating in science. These 'manifold costs' include spending a greater amount of effort and time reading, writing, and disseminating work (Amano et al., 2023). Even when NNES do not use GenAI tools, research suggests that they are likely to suffer false accusations. Liang et al (2023) found that GenAI text detectors falsely detect NNES writers' outputs as AI-Generated due to the higher level of perplexity and lower coherence often found in NNES writing. As a result, linguistic measures, such as perplexity, as indicators of GenAI, are associated with bias against NNES (Liang et al., 2023). Following this, although GenAI detectors are borne from a place of good intention, that is, to limit the ability of individuals to commit AI-enabled academic integrity violations, to understand how students use GenAI, and to ensure fairness and parity in assessment submissions, thus promoting inclusivity and fairness, they have unintended, paradoxical consequences. Most concerning among these is that they can be used to generate false accusations of academic dishonesty, create confusion or heightened anxiety among students, and disadvantage some groups of students (e.g. NNES) in comparison to others.

To assess whether the benefits GenAI text detectors provide to the educational process (ensuring fairness by limiting the scope of AI-enabled authorship misrepresentation) outweigh their potential to drive inequalities (false accusations of NNES, potential for students with technological ability, and financial resources to bypass detection systems), we investigate the efficacy of six GenAI text detectors. We do this by testing content subjected to prompting techniques that alter the standard output of GenAI tools to disguise that they are AI-generated texts. We refer to these manipulation techniques designed to evade detection as 'adversarial techniques' in reference to the combination of methods that have been referred to as adversarial attacks in the extant literature related to attacks on machine learning based tools (Alsmadi et al., 2021; Qiu et al., 2019; Sadasivan et al., 2023; Wang et al., 2023). This contributes to the understanding of the limitations of such AI text detector tools in light of their risks to inclusivity in educational practice, given that prior research has not investigated the impact of adversarial techniques on the ability of AI text detectors to maintain a reliable detection accuracy rate.

## Objectives

The overarching aim of this research is not to vilify the use of GenAI tools or AI text detectors, but to explore the efficacy and sensitivity of these tools to linguistic changes and adversarial prompting techniques, thereby assessing their ability or inability to foster an inclusive system of education. By 'inclusivity' we refer to providing equal opportunities to all who participate in the educational process and ensuring that no group of students is at a disadvantage through a lack of opportunity or unfair assessment process. By measuring the susceptibility of existing AI text detectors to various adversarial techniques and assessing their efficacy, we can draw conclusions about their suitability for use in higher education and their ability to enable inclusive practices by reducing the risk of academic misconduct or detracting from inclusivity through unreliability (thus leading to a higher likelihood of false positives). In doing so, this research aims to serve as a guide for educators and institutions regarding the use of AI text detectors.

To achieve these objectives, this study explores whether AI text detectors can be considered reliable tools which contribute to academic integrity and offer a net positive rather than negative (given their bias) in fostering inclusivity and fairness. Specifically, we aim to answer the following questions.

1. **Reliability of AI Text Detectors**: To what extent can we trust the results of AI text detectors to accurately determine the source of a piece of text? What implications does this have for inclusive assessment practices?
2. **Adversarial Techniques**: Is it possible to disguise AI generated content using adversarial techniques in prompting strategies and paraphrasing? What are the most effective adversarial techniques for deceiving text detectors? How might this advantage some writers over others?
3. **Comparative Analysis of AI Tools**: Which AI tools provide outputs that are easier or more challenging for detectors to identify? What are the implications of this?
4. **Detector Efficacy**: Which AI text detectors show more promise in terms of accuracy and reliability? Is it possible to recommend any of them based on their performance?

## Literature

### Generative AI Tools

Educators and researchers are currently grappling with the changes brought about by the popularisation and development of Foundation Models (FMs). These models share the fundamental capability of generating human-like texts in response to natural language prompts, a feature that enables a wide range of applications, and this study focuses on the outputs created by three popular FMs: GPT-4 by OpenAI (accessed through the ChatGPT interface), Bard (now known as Gemini) by Google, and Claude 2 by Anthropic.

In an evaluation of GPT-4, Claude 2, and Bard's performance, Borji and Mohammadian (2023) identified GPT-4 as providing correct answers to standardized questions 84% of the time, while Claude 2 and Bard achieved scores of 64.5% and 62.4%, respectively. In the multiple-choice section of the Bar exam, Claude 2 excelled with a 76.5% accuracy rate, surpassing GPT-4's previous record of 75%. For tasks involving multistep reasoning, Fu et al. (2022) concluded that Claude 2's performance is comparable to that of GPT-4, with Lin and Chen (2023), claiming that Claude 2 slightly outperforms GPT-4. Despite these achievements, Claude 2 lags in scientific writing and quantitative accuracy compared to other models (Chang et al., 2023; Lozić & Štular, 2023; Z. Wu et al., 2023), primarily because of the ability of GPT-4 and Bard to access the internet for information gathering. Studies by Fu et al. (2022) and Lin and Chen (2023) have shown that Claude 2 can perform comparably to or even slightly outperform GPT-4 in multistep reasoning and dialogue.

### Efficacy of AI Text Detection Tools

There is a wide range of software available which has been designed to classify whether text is machine or human generated, with providers claiming high levels of accuracy in being able to identify whether text is written by a human or by a GenAI tool (GPTZero, n.d.; Turnitin, 2023). While some of these tools are free and others require either registration or payment, research by Walters (2023) has identified that the accuracy of paid-for tools is only slightly higher than that of free versions. However, claims of accuracy are contradicted by studies which demonstrate the varied levels of the detectors' ability to distinguish accurately between AI and human-generated content. (Chaka, 2023a; Gao et al., 2022; Krishna et al., 2023; Orenstrakh et al., 2023; Perkins, Roe, et al., 2023; Walters, 2023; Weber-Wulff et al., 2023).

Detection tool biases against NNES have been highlighted as a potential issue of bias by Fröhling and Zubiaga (2021) and Liang et al. (2023) because of their reliance on standardised linguistic metrics such as perplexity and burstiness, which can disadvantage NNES. Liang et al. (2023), identified that GPT-based detection tools misclassified over half of the samples from NNES, with an average false positive rate of 61.3%. However, research produced by GPTZero challenges Liang's findings and demonstrate that GPTZero can accurately determine the human written status of text when produced by an NNES (Tian, 2023). This claim is repeated in research produced by Turnitin who also show no differences in how NNES writing is classified (Adamson, 2023). However, on a broader scale, questions remain, especially as OpenAI's AI classifier, although trained on varied human textual patterns, did not include training on NNES-generated text (Elkhatat et al., 2023; OpenAI, 2023). Owing to the inability of this classifier to accurately detect the output of GenAI tools, OpenAI withdrew this tool from use in July 2023 (OpenAI, 2023). Research by Originality.AI (2023) acknowledges that all AI text detectors have their limitations. These include lagging behind in training against recent GenAI tools, which have reached a level of complexity in producing content, making it very difficult to distinguish between AI and human-generated content, and that they are susceptible to being bypassed by using adversarial techniques. These problems underscore the need for educational institutions to balance the use of AI detection tools with accommodating AI-produced materials (Perkins, Roe, et al., 2023).

**Adversarial techniques as a method of evading GenAI detection**

A few previous studies have investigated how the accuracy of AI text detectors can decrease following manipulation of the textual output of GenAI tools. Mitchell et al. (2023) identified that DetectGPT correctly detected 70.3% of model-generated sequences from GPT2-XL. However, after altering the content using an Automated Paraphrasing tool (APT), the detection rate decreased to 4.6%. Similar results of major drops in accuracy rates following the application of APTs and translation tools were identified by Weber-Wulff et al. (2023).

A key technique for reducing detection accuracy involves the deliberate incorporation of errors in AI-generated text weakness (Perkins, Roe, et al., 2023). This technique exploits the natural occurrence of minor mistakes in human writing, such as typographical errors, grammatical inconsistencies, and stylistic irregularities. By mimicking these imperfections, AI-generated content can effectively mislead detectors into classifying them as human-authored content. Another approach, as discussed by Liang et al. (2023), involves tailoring the complexity of a text to mirror the style typically found in human-produced academic works. This method focuses on adjusting language use and vocabulary to align closely with human writing styles. Thus, AI-generated content can blend more seamlessly with human-authored texts, thereby evading detection more effectively.

The role of prompting GenAI models ('prompt engineering') has also been emphasised as important when evading the detection of machine-generated text. As observed in studies by Elkhatat et al. (2023), carefully crafted prompts can guide GenAI tools in generating text that not only aligns with the desired content but also strategically incorporates elements that challenge the detection capabilities of AI detectors. This involves a nuanced understanding of the capabilities of AI and the limitations of detectors, ensuring that the generated content remains undetectable while retaining its intended meaning and coherence. Other techniques for evading the identification of machine-generated text have also been presented by Solaiman et al. (2019), Sadasivan et al. (2023) and Ippolito et al. (2020). These include "recursive paraphrasing attacks", which utilize automated, network-based paraphrasing tools to reduce the accuracy of detectors on watermarked texts, and "spoofing attacks" in which an adversarial human deliberately writes a passage falsely detected as AI-generated without having access to the inner workings of the detection methods. Lancaster (2023) proposed that watermarking AI-generated text may be a potential solution; however, the attacks discussed above demonstrate that even LLMs protected by watermarking schemes are vulnerable to manipulation of this type.

**Research gap**

Despite the growing interest and concern surrounding the reliability of AI text detectors and the effectiveness of adversarial techniques in deceiving them, there are no comprehensive comparative analyses which systematically evaluate the performance of various AI text detectors against different GenAI tools, nor are there any studies that do this while exploring the potential consequences from an inclusivity and equity perspective. Existing research points to a lack of holistic understanding of the interplay between AI text detectors, GenAI models, and adversarial techniques (Anderson et al., 2023; Chaka, 2023b; Elali & Rachid, 2023; Elkhatat et al., 2023; Liang et al., 2023; Orenstrakh et al., 2023; Perkins, Roe, et al., 2023; Weber-Wulff et al., 2023). This research gap impedes our ability to provide robust evidence-based recommendations for educators regarding the use of AI text detectors. Gaining a more nuanced perspective on how accurate AI text detectors are in real-world settings means that we can highlight their potential benefits, limitations, and biases and therefore support a more inclusive educational environment in this new GenAI-infused era.

To support this goal, this study explores the effectiveness of different adversarial techniques in adjusting the output of GenAI tools to evade detection by AI text detectors. In doing so, we address the need for a standardised framework and taxonomy to compare and evaluate the performance of AI text detectors against GenAI tools and adversarial techniques (Abd-Elaal et al., 2022). With a growing number of GenAI tools available, each having unique characteristics, strengths, and weaknesses, students attempting to evade detection may employ various techniques to manipulate AI-generated text, making it less likely to be identified by text detectors. Understanding these manipulation techniques can help demonstrate the true effectiveness of AI text detectors in more realistic settings, where students actively try to avoid detection. This knowledge can inform recommendations for higher education institutions and academics regarding whether and how to effectively use these tools to support academic integrity. In addition, these insights can guide software developers in creating more robust detection methods to counter evasion techniques and further enhance the reliability of AI text detectors.

# Methodology

## Overview

This study employs an experimental design in which we use three popular GenAI tools to generate short samples of text (n=15). Altered versions of the original samples are created by applying six adversarial techniques (n=89). Ten human-written samples are used as controls. All the developed samples (n=114) are tested against seven popular AI text detectors to determine the effect of adversarial techniques on the accuracy of AI text detectors (n=805). Sample creation and testing were conducted in September and October 2023.

All prompts used to generate the samples, as well as the samples themselves are publicly available at Mendeley Data at the following link https://data.mendeley.com/datasets/xv6fk2mmh9/2 (Perkins et al., 2024).

## GenAI Sample Generation

First, we generated an initial set of 15 text samples using three GenAI tools (GPT4, Claude 2, and Bard), with five samples created using each tool. All samples were created using the same five prompts designed to emulate a range of tasks in which GenAI tools may be used. These prompts requested the development of the following outputs.

- Mini (short form) university essay testing AI's ability to construct coherent, argumentative, or exploratory work within an HE setting.
- Professional blog post to assess whether GenAI-generated content can demonstrate professionalism, industry knowledge, and expertise, while maintaining reader engagement.
- Cover letter to apply for an internship designed to test GenAI's ability to design tailored content specific to a position and the suitability and motivation of the applicant.
- Middle-school level comparative analysis task designed to test GenAI's use of language specific to that of a younger author requiring clarity and simplicity.
- Magazine article intended to test for content and tone in a journalistic manner that may appeal to a broad audience.

The prompts were designed to create outputs which covered a range of different topics, linguistic styles, and complexity levels to provide insights into the versatility and consistency of AI detectors in producing results across different forms of text.

Ten control samples were created by the human authors (researchers in this project). The human authors are all proficient English speakers, comprising academic staff who described English as their primary language, and Vietnamese undergraduate NNES students who study at a UK-oriented international university in English.

## Adversarial sample generation

All original 15 AI-generated samples were modified using six different adversarial techniques. With the exception of a separate APT tool to paraphrase the outputs, all adversarial techniques were applied using the same FM with which the samples were originally generated. This maintains the likely process of a user interacting with their chosen FM. This was performed by copying the sample to a new chat window in each tool and then using prompting strategies to request the text to be adjusted according to the requirements of each adversarial technique.

All adversarial techniques were applied following a standardised protocol. The testers were given a standard prompt format for application to each of the 15 samples. These prompts include specific techniques designed to improve the performance of GenAI tools. Examples include the use of delimiters (Sweenor & Ramanathan, 2023) and prompt chaining (T. Wu et al., 2022). Delimiters (in this case, hashmarks and titles) were used to identify to ChatGPT where instructions ended, and the content for adjustment began. If a suitable output was not obtained on the first attempt, prompt chaining was used to request specific adjustments to meet the required output. Researchers were restricted to making up to five requests for new prompts to obtain suitable outputs. If a suitable output could not be created within these five iterations, the sample was not considered for testing. Any potential ambiguities in prompts were addressed by requesting the GenAI tool to ask any required questions before providing output. At no point were the text samples manually edited: all adjustments to the original samples were carried out using the relevant GenAI tool.

The format used for each adversarial technique is presented in Table 1.

| Technique | Prompt format |
|---|---|
| Add Spelling Errors (SE) | **Topic**: "Rewriting with Intentional Errors"<br>**Instructions**: Rewrite the following text passage with spelling errors.<br>**Style**: someone who is not proficient in English spelling. However, the errors should not be so extreme that the text becomes incomprehensible. Aim for errors that are commonly seen in writing by individuals who have a good grasp of the English language#<br>**Text Passage for Rewriting: [Insert text here]**#<br>**Word Count**: approximately 500 #<br>Ask any questions that you need clarifying before producing the output |
| Increase Burstiness (IB) | **Topic**: "Varying Sentence Length "<br>**Instructions:** Rewrite the following passage with the aim of varying sentence length to create a more dynamic and engaging text. Use a mix of short, medium, and long sentences to achieve this effect. Adjust paragraphs so that these are also of a different length#<br>**Text Passage for Rewriting: [Insert text here]**#<br>**Note:** While varying sentence length, ensure that the text remains coherent and academically appropriate. The goal is to make the writing more engaging and human sounding without sacrificing its core meaning.#<br>**Word Count:** approximately 500 |
| Increase Complexity (IC) | **Topic**: "Increasing Text Complexity "<br>**Instructions:** Rewrite the following passage to significantly enhance its linguistic complexity. Use specialized vocabulary, intricate sentence structures, and nuanced arguments to make the text more advanced.#<br>**Text Passage for Rewriting: [Insert text here]**#<br>**Note:** While increasing the complexity, ensure that the text remains coherent and that the core arguments are not lost. Feel free to incorporate academic jargon or technical terms that are relevant to the subject matter.#<br>**Word Count:** approximately 500#<br>Ask any questions that you need clarifying before producing the output |
| Decrease Complexity (DC) | **Topic**: "Downgrading Text Complexity<br>**Instructions:** Rewrite the following passage to reduce the overall complexity, making it simpler to understand.<br>**Text Passage for Rewriting: [Insert text here]**#<br>**Note:** The rewritten text should maintain the essential points but use simpler vocabulary and sentence structures. The aim is to produce a text that is less complex and linguistically sophisticated than the original output.# |
| | **Word Count:** approximately 500#<br>Ask any questions that you need clarifying before producing the output |
| Write as NNES<br>NNES) | **Topic**: "Rewriting as a Non-Native English Speaker (NNES) with IELTS Band Level 6"<br>**Instructions**:<br>Rewrite the following text passage to reflect the writing style of a non-native English speaker who has achieved a band level 6 in IELTS writing. This level indicates a competent user of English, but with some inaccuracies, inappropriate usage, and misunderstandings. The text should be mostly clear but may contain occasional errors in grammar, vocabulary, and coherence.#<br>**Text Passage for Rewriting**: [**Insert text here**]#<br>**Note**: Aim for errors that are typical of an IELTS band level 6 writer. These could include minor grammatical mistakes, slight misuse of vocabulary, and occasional awkward phrasing. However, the overall meaning of the text should remain clear and understandable.#<br>**Word Count**: approximately 500#<br>Ask any questions that you need clarifying before producing the output |
| Paraphrase<br>(PR) | Paraphrasing was carried out using the free version of the APT 'Quillbot' (Paraphrasing Tool - QuillBot AI, n.d.) in the standard mode. |

*Table 1. Adversarial technique prompts*

Suitable outputs could not be obtained in one case, resulting in 89 samples of AI-generated text in which adversarial techniques had been applied. When creating the samples for testing, we considered using few-shot prompting techniques to guide the AI models in generating more human-like text. However, we ultimately chose not to employ this method to minimise the influence of external text on the final output. By allowing the AI tools to apply adversarial techniques based on their own underlying datasets and processes, we aimed to demonstrate how each tool would uniquely interpret and execute these techniques, thereby providing a more authentic representation of their capabilities. Table 2 highlights key points related to the application of each technique.

| Technique | Prompt format |
|---|---|
| Add Spelling Errors<br><br>(SE) | This technique was designed to increase the number of spelling errors found in the output of a text to mimic the work of someone who had not carefully proof-read their work. This prompt resulted in outputs which generally contained a minimum of 20 errors, even when prompt chaining requested a reduced number of errors. Following discussions among authors it was decided that samples with a larger number of errors would be accepted for testing. |
| Increase Burstiness<br><br>(IB) | This technique is designed increase the 'burstiness' or variability in the sentence lengths of responses, a key determinant used by some AI text detectors such as GPTZero (Tian & Cui, 2023) to identify AI generated text. By including a range of sentences varying in length and structure, the aim was to create sentences with varied lengths that reflected a more typical human writing pattern. The outcome was text that frequently sounded more human-authored, but sometimes resulted in overly short sentences that might not be suitable for formal or professional contexts. |
| Increase Complexity<br><br>(IC) | This technique was designed to produce text that increases the complexity of writing styles by adding jargon, increasing sentence length, and using more convoluted grammatical structures. Following the application of this technique, significant changes were made to the output, albeit at the expense of readability and stylistic consistency. We observed that the outputs often descended into jargon or strayed significantly from the expected style which may result in suspicion as to the authorship of the work, similar to the 'word salad' that frequently appears following the use of an APT (Roe & Perkins, 2022; Rogerson & McCarthy, 2017). |
| Decrease Complexity | This technique requested a simplification of a sample's vocabulary and sentence structures, with the goal being to reduce the linguistic complexity while preserving the core message of the sample, and therefore potentially highlighting the authorship of a text. |
| (DC) | |
| Write as NNES (NNES) | This technique aimed at mimicking the writing style of a NNES. In some ways the concept of a 'NNES' is problematic, as it may lead to a stereotypical output which reflects societal and cultural biases in the training data. This adversarial method sought to generate text embodying certain inaccuracies, inappropriate usage, and misunderstandings typical of a NNES possessing a competent yet not advanced level of English proficiency. However, it was decided that any output produced using this technique needed to be comprehensible despite any errors. |
| Paraphrase (PR) | This technique was designed to test the ability of existing APT software tools in misleading AI text detectors by rewriting the original material using different lexis, phrasing, and sentence construction to express the same ideas. This reflects typical strategies that can be used to pass traditional plagiarism checking software without adjusting the main message of the original sample. |

*Table 2. Notes regarding adversarial techniques*

**Summary of sample generation**

Owing to limitations inherent in some of the AI text detectors used, the length of the samples requested in all cases was restricted to approximately 500 words. This resulted in the final length of the samples for testing falling between 350 and 625 words.

Table 3 summarises the different samples generated.

| Sample description | Number of samples created |
|---|---|
| Human written control samples (written by authors) | 10 |
| AI generated samples (developed from five standardised prompts) | 15 |
| AI generated samples with adversarial techniques applied (Six techniques applied in a standardised manner to each of the 15 samples) | 89 |
| **Total** | **114** |

*Table 3. List of samples for testing*

**Testing protocol**

All samples created were tested against seven AI text detectors which had been previously identified in literature (Chaka, 2023a; Gao et al., 2022; Krishna et al., 2023; Orenstrakh et al., 2023; Perkins, Roe, et al., 2023; Weber-Wulff et al., 2023)as being somewhat effective at successfully identifying GenAI content. The detectors were chosen based on a review of existing studies (Chaka, 2023a; Gao et al., 2022; Krishna et al., 2023; Orenstrakh et al., 2023; Perkins, Roe, et al., 2023a; Weber-Wulff et al., 2023) as follows:

- Turnitin AI detect (Turnitin.com, 2023)
- GPTZero (Tian & Cui, 2023)
- ZeroGPT (ZeroGPT.com, n.d.)
- Copyleaks (Copyleaks.com, n.d.)
- Crossplag (Crossplag.com, n.d.)
- GPT-2 Output Detector (OpenAI, n.d.)

- GPTKit (GPTKit.com, n.d.)

To test the samples, the text was either copied and pasted onto online software tools or uploaded where this option was available. The tools used free versions of the software, where possible. However, Turnitin's AI detection software was used as part of the licence held by the lead author's institution, and both GPTKit and Copyleaks required credits to be purchased to run the required number of samples. All 114 samples were tested using each of the six detection tools, resulting in a total number of 805 tests. Eight samples could not be tested using some of the tools, resulting in a total of 797 valid tests.

The results provided by the AI detection tools were recorded and interpreted based on a modified version of the protocol designed by Weber-Wulff et al. (2023) for comparison with the existing literature. All test results were classified on a scale as either Positive (P) or Negative (N) based on the accuracy of the tool in determining whether the output was written by a human or generated by a GenAI tool. The classification is presented in Table 4.

| Human-written (NEGATIVE) and the tool says that it is written by: | | |
|---|---|---|
| [100 - 80%) human/ [0-20%] AI | TN | True negative |
| [80 - 60%) human/ [20-40%] AI | PTN | Partially true negative |
| [60 - 40%) human/ [40-60%] AI | UNC | Unclear |
| [40 - 20%) human/ [60-80%] AI | PFP | Partially false positive |
| [20 - 0%] human/ [80-100%] AI | FP | False positive |
| AI-written (POSITIVE) and the tool says that it is written by: | | |
| [100 - 80%) human/ [0-20%] AI | FN | False negative |
| [80 - 60%) human/ [20-40%] AI | PFN | Partially false negative |
| [60 - 40%) human/ [40-60%] AI | UNC | Unclear |
| [40 - 20%) human/ [60-80%] AI | PTP | Partially true positive |
| [20 - 0%] human/ [80-100%] AI | TP | True positive |

*Table 4. Classification scale*

**Interpretation protocol**

As the AI text detectors all display their results differently, with some providing quantitative statements ("X% probability for AI") and some qualitative statements ("most likely human written, we needed a way to standardise the output to allow for comparability. Therefore, we recorded the results based on the interpretation protocols shown in Table 5.

| Tool | Result | Human-written | AI-written |
|---|---|---|---|
| **Turnitin** | "… [0-20%] of the text is generated by AI" | TN | FN |
| | "… [20-40%] of the text is generated by AI" | PTN | PFN |
| | "… [40-60%] of the text is generated by AI" | UNC | UNC |

| Tool | Result | Human-written | AI-written |
|------|--------|---------------|------------|
|  | "… [60-80%] of the text is generated by AI" | PFP | PTP |
|  | "… [80-100%] of the text is generated by AI" | FP | TP |
| **GPT Zero** | "likely to be written entirely by human" | TN | FN |
|  | "may include parts written by AI" | UNC | UNC |
|  | "likely to be written entirely by AI" | FP | TP |
| **ZeroGPT** | "… is Human written" | TN | FN |
|  | "… most likely Human written" | TN | FN |
|  | "… most likely Human written, may include parts generated by AI/GPT" | PTN | PFN |
|  | "... likely Human written, may include parts generated by AI/GPT" | PTN | PFN |
|  | "… contains mixed signals, with some parts generated by AI/GPT" | UNC | UNC |
|  | "… is likely generated by AI/GPT" | PFP | PTP |
|  | "… is most likely AI/GPT generated" | PFP | PTP |
|  | "… most of Your text is AI/GPT generated" | FP | TP |
|  | "… is AI/GPT generated" | FP | TP |
| **Copyleaks** | "… [0-20%] probability for AI" | TN | FN |
|  | "… [20-40%) probability for AI" | PTN | PFN |
|  | "… [40-60%) probability for AI" | UNC | UNC |
|  | "… [60-80%) probability for AI" | PFP | PTP |
|  | "… [80-100%] probability for AI" | FP | TP |
| **Crossplags** | "… mostly written by human" | TN | FN |
|  | "… may include parts written by AI" | UNC | UNC |
|  | "… mostly written by AI" | FP | TP |
| **GPT-2 output detector** | "… [80-100%] real" | TN | FN |
|  | "… [60-80%) real" | PTN | PFN |
|  | "… [40-60%) real" | UNC | UNC |
|  | "… [20-40%) real" | PFP | PTP |
|  | "…[0-20%] real" | FP | TP |
| **GPTKit** | "… [80-100%] real" | TN | FN |
|  | "… [60-80%) real" | PTN | PFN |
|  | "… [40-60%) real" | UNC | UNC |
|  | "… [20-40%) real" | PFP | PTP |
|  | "…[0-20%] real" | FP | TP |

*Table 5. Interpretation protocol for AI detectors*

**Accuracy and error analysis**

After classifying the outcomes of the tools as (partially) true/false positives/negatives, we assessed the accuracy of the tools using three separate methods, as identified by Weber-Wulff et al. (2023). These included a binary classification in which the accuracy was calculated as the ratio of correctly identified cases to all cases, a semi-binary approach which allowed for the results of partially correct results to be awarded half scores, and a logarithmic approach in which scores increased as accuracy rates increased. We also report the mean values of the combined methods.

The calculations used for each test are presented in Table 6.

| Accuracy method | Calculation |
|---|---|
| Binary | $Acc\_Bin = (TN + TP) / (TN + PTN + TP + PTP + FN + PFN + FP + PFP + UNC)$ |
| Semi-binary | $Acc\_SemiBin = ((TN + TP) + 0.5 * (PTN + PTP)) / (TN + PTN + TP + PTP + FN + PFN + FP + PFP + UNC)$ |
| Logarithmic | $Acc\_Logarithmic = (1 * (FN + FP) + 2 * (PFN + PFP) + 4 * UNC + 8 * (PTP + PTN) + 16 * (TP + TN)) / (TN + PTN + TP + PTP + FN + PFN + FP + PFP + UNC)$ |
| Average | $Acc\_Final = (Acc\_Bin + Acc\_SemiBin + Acc\_Logarithmic) / 3$ |

*Table 6. Calculations of accuracy rates*

In addition, error analysis was conducted to assess false accusations and undetected cases, which are particularly relevant in ensuring an inclusive and non-discriminatory environment. Instances of AI detectors falsely accusing students of academic misconduct are not uncommon and cause concern regarding inclusivity, fairness, and ethical practice in education. Therefore, it is important to establish a threshold at which the AI-generated content detected by these tools is sufficient to level an accusation or to take disciplinary actions. For this calculation, we used a threshold value of 60% to represent a value higher than chance, and to recognise the low likelihood of a partially negative or unclear value resulting in a false accusation of a student (FAS):

$FAS = (FP + PFP) / (TN + PTN + TP + PTP + FN + PFN + FP + PFP + UNC)$

Another type of error, which is the potential for AI detectors to fail to identify AI-produced texts, can lead to students who use AI for unauthorised content generation being awarded similar credit as honest ones. The likelihood of undetected cases (UDC) is calculated as follows:

$UDC = (FN + PFN) / (TN + PTN + TP + PTP + FN + PFN + FP + PFP + UNC)$

## Results

### Baseline testing

The results of our baseline testing of 15 AI-generated samples and 10 human samples are shown in Table 7. This test aims to establish the abilities of AI detection tools to determine the authorship criteria of a given sample before any adversarial techniques are applied.

| Rank | BY AI DETECTORS | MEAN | ACCURACY | | |
|---|---|---|---|---|---|
| | | | Binary | Semi-binary | Logarithmic |
| 1 | Copyleaks | 64.8% | 64% | 64% | 66.3% |
| 2 | Turnitin | 61% | 56% | 62% | 65% |
| 3 | Crossplag | 60.8% | 60% | 60% | 62.5% |
| 4 | GPT-2 detector | 57.2% | 56% | 56% | 59.5% |
| 5 | ZeroGPT | 46.1% | 40% | 460% | 52.3% |

| Rank | BY AI DETECTORS | MEAN | ACCURACY | | |
|---|---|---|---|---|---|
| | | | Binary | Semi-binary | Logarithmic |
| 6 | GPTKit | 37.3% | 32% | 36% | 43.8% |
| 7 | GPTZero | 26.3% | 16% | 24% | 39% |
| BY GENAI TOOLS | | | | | |
| 1 | Bard | 76.9% | 71.4% | 78.6% | 80.7% |
| 2 | GPT-4 | 23.9% | 20% | 21.4% | 30,2% |
| 3 | Claude 2 | 17.7% | 14.3% | 15.7% | 23.2% |
| OVERALL | | | | | |
| Original AI-generated samples | | 39.5% | 35.2% | 38.6% | 44.7% |
| Control samples | | 67% | 62.9% | 66.4% | 71.6% |

*Table 7. Baseline testing*

We wish to highlight that these results were accurate at the time of testing, but because of the evolution of both the FMs and AI text detection tools, these comparative results may not necessarily represent the current ability of any of the listed tools.

Testing shows that the AI detectors tested have a mean accuracy rating of only 39.5% when evaluating unmanipulated AI-generated content. Importantly, regarding the human-written control samples, only 67% of the tests were accurate, leading to significant concerns regarding the potential for false accusations from these tools.

Regarding the performance of individual AI detectors, Copyleaks displayed the highest sensitivity, being able to detect 64,8% of AI-generated texts, followed closely by Turnitin at 61%. The worst-performing detector was GPTZero, with a considerably lower accuracy rating of approximately 26%. Before the content was subjected to any manipulation, the GenAI tool which produced the most easily detectable text was Bard, with 76.9% of its outputs correctly identified as AI generated by the detector tools.

**Accuracy of AI detectors following the application of adversarial techniques**

Next, the performance of the AI detectors was assessed based on their average accuracy rate when detecting manipulated outputs. Table 8 shows the accuracy of the tools when faced with non-manipulated AI content, the accuracy following the application of adversarial techniques, and percentage drops in accuracy. All reported figures are the mean values.

| Ranking | AI Text Detector | Accuracy (non-manipulated output) | Accuracy (manipulated output) | % drop in accuracy |
|---|---|---|---|---|
| 1 | Copyleaks | 73.9% | 58.7% | 15.2% |
| 2 | Crossplags | 54.3% | 32.4% | 21.9% |
| 3 | GPT-2 output | 34.7% | 17.5% | 17.2% |
| 4 | ZeroGPT | 31.3% | 17.3% | 14% |
| 5 | GPTZero | 26.4% | 16.7% | 9.7% |
| 6 | Turnitin | 50% | 7.9% | 42.1% |
| 7 | GPT Kit | 6% | 4.5% | 1.5% |
| **Average** | | **39.5%** | **22.2%** | **17.4%** |

*Table 8. Accuracy rate as an average of 3 approaches discussed, by AI detectors*

A comparison of the tools showed reductions in accuracy with variations ranging from 1.5% to more than 42% when the outputs were subjected to adversarial techniques (mean value 17.4%). Copyleaks was the most accurate (58.7%), whereas GPT Kit was the least accurate (4.5%).

Turnitin, one of the most widely used platforms in HEIs, demonstrated the highest drop in accuracy (42.1%) when testing the manipulated output. As a result, despite having the second-highest accuracy in baseline testing, this detector only ranked 5th place out of the seven AI detectors in terms of accuracy following the application of adversarial techniques.

The effectiveness of each technique in reducing the ability of the AI-generated text to be detected is shown in Table 9, ranked by the overall percentage drop in accuracy after the application of each technique.

| Ranking | Adversarial Technique | Accuracy | % drop in accuracy |
|---|---|---|---|
| 1 | Add Spelling Errors (SE) | 12.9% | 27% |
| 2 | Increase Burstiness (IB) | 15.9% | 24% |
| 3 | Paraphrase (PR) | 18.4% | 21% |
| 4 | Decrease Complexity (DC) | 21% | 19% |
| 5 | Write as NNES (NNES) | 27.7% | 12% |
| 6 | Increase Complexity (IC) | 37% | 2% |
| **Mean** | | **22.1%** | **17.5%** |

*Table 9. Accuracy rate as an average of 3 approaches discussed, by adversarial techniques*

Regarding techniques which have a major impact on reducing text detectability, outputs produced with spelling errors (12.9%) or higher burstiness (15.9%) were almost undetectable. However, we recognise that an examination of many of the samples produced using the SE technique resulted in an output that would be very unlikely to be submitted by a student. Although they evaded detection, they would very likely receive poor marks in a real-world setting because of the high number of errors. Increasing the complexity of texts was the least effective adversarial technique, with only a marginal drop of 2% in the detectability rate. Overall, the application of adversarial techniques resulted in a 17.5% drop in accuracy compared to non-manipulated content.

**Error analysis**

Table 10 presents an analysis of the errors produced by AI detectors with a focus on false accusations against human-written samples and the proportion of undetected machine-generated samples.

| AI Text detector | False accusation | Undetected cases | FAS | UDC |
|---|---|---|---|---|
| Turnitin | 0 | 88 | 0% | 84% |
| GPT Zero | 1 | 58 | 10% | 55% |
| Zero GPT | 0 | 77 | 0% | 73% |
| Copyleaks | 5 | 41 | 50% | 39% |
| Crossplag | 3 | 69 | 30% | 66% |
| GPT-2 output | 0 | 81 | 0% | 77% |
| GPT Kit | 0 | 89 | 0% | 85% |
| **Total/Mean** | **9** | **414** | **15%** | **65.7%** |

*Table 10: FAS and UDC ratios*

Considerable disparities were observed across detectors in terms of false accusations. However, four of the seven detectors did not misclassify any of the human-written samples. Notably, despite detecting the highest proportion of manipulated text, Copyleaks possessed the highest likelihood of producing false results, with 50% of human-generated samples misidentified as AI written. Regarding undetected cases, Table 10 also reveals a major risk to academic integrity from the use of adversarial techniques, with six out of seven detectors having a UDC ratio exceeding 50%, and even the highest-performing tool (Copyleaks) failing to identify 39% of AI-generated cases.

**Comparative Performance of Generative AI tools**

As discussed earlier, the GenAI tools (Bard, Claude 2, GPT-4) chosen to generate the samples and apply adversarial techniques possess inherent differences in their capabilities, which stem from the different architectures used, and sizes of the training datasets employed. Table 11 highlights some noticeable features regarding the differences between these tools and their potential impact on how well they were able to reduce the detectability of the original samples.

| Adversarial Technique | Bard | | Claude 2 | | GPT-4 | |
|---|---|---|---|---|---|---|
| | Accuracy | % reduction | Accuracy | % reduction | Accuracy | % reduction |
| Original AI samples | 76.9% | - | 17.7% | - | 23.9% | - |
| Add spelling errors (SE) | 14.3% | 62.6% | 7.9% | 9.9% | 16.5% | 7.3% |
| Write as NNES (NNES) | 16.4% | 60.5% | 2.2% | 15.5% | 29.0% | -5.2% |
| Decrease complexity (DC) | 57.2% | 19.7% | 14.6% | 3.2% | 34.9% | -11.0% |
| Increase complexity (IC) | 50.4% | 26.5% | 10.8% | 6.9% | 2.1% | 21.7% |
| Increase burstiness (IB) | 57.7% | 19.2% | 14.1% | 3.6% | 11.3% | 12.6% |
| Paraphrase (PR) | 32.4% | 44.5% | 8.8% | 8.9% | 13.9% | 10.0% |
| Mean accuracy reduction | - | 38.8% | - | 8.0% | - | 7.6% |

*Table 11. Performance of Generative AI tools*

Overall, we observed that the reduction in detectability was highest when applying adversarial techniques using Bard, with a mean reduction in accuracy of 38.8%. Claude 2 and GPT-4 ranked very closely, at 8% and 7.6%, respectively. However, this must be considered in light of the high accuracy of the AI detectors in correctly identifying the output from Bard in the original AI-generated samples. Therefore, in these cases, adversarial techniques had an oversized impact on reducing the detectability of the Bard output. It is also notable that the three GenAI tools exhibited unique patterns in performance when applying different adversarial attacks. For example, outputs from Bard and Claude 2 showed reductions in detectability when all techniques were applied, whereas applying certain techniques in GPT-4 resulted in an output that was easier to detect. This suggests that the choice of the GenAI tool has an impact in which adversarial techniques might be chosen if reducing the detectability of the text is the goal.

In comparison to Weber-Wulff (2023) et al.'s testing of detection tools, we noted that AI detectors performed significantly worse in our study, even with unedited text. In their analysis, all detectors achieved accuracy rates of less than 80%, but five scored above 70%. However, in our baseline testing, no detectors achieved such high accuracy. Furthermore, while Weber-Wulff et al. (2023) found that Turnitin's detector was the most accurate, our results placed Turnitin at number two, with Copyleaks showing greater accuracy. The application of machine translation in their study led to a 20% reduction in accuracy, whereas human manual editing reduced accuracy to approximately 50%, and machine translation (paraphrase) reduced the overall accuracy to 26%. Our results showed similar impacts, with paraphrasing as a technique that reduced accuracy by 21%. In reference to the findings of Elkhatat et al. (2023) and Perkins et al. (2023), we also noted a high degree of variability and low consistency when dealing with GPT-4 content. Our findings also corroborate Chaka's (2023) and Walters' (2023) research, demonstrating that Copyleaks appears to be the most accurate of the current generation of GenAI text detectors (despite a high FAS rate), with Turnitin also performing relatively well compared to other detectors. These results contribute to the body of knowledge about the variable accuracy of these tools and highlight the potential for inequity in educational assessments across cohorts if detection software is used.

## Discussion

### Accuracy and vulnerabilities of AI text detectors as a barrier to inclusivity

In our baseline testing protocol of both non-manipulated AI-generated samples tested alongside the human-written control samples, we see an initially lower than expected average accuracy rating for the detection of AI-generated content, coupled with a substantial rate of false accusations in the human-written control samples. When the AI-generated samples were subjected to manipulation, significant vulnerabilities in accurately detecting text were observed. If the goal of implementing AI detection tools as part of an overall academic integrity strategy is to support academic staff in identifying where machine-generated content has been used and has not been declared, these inaccuracies may lead to a false sense of security and a broader reduction in assessment security. As assessment security is a key component in ensuring inclusive, equitable, and fair opportunities for learners, this is problematic. The varying degrees of reduction in accuracy following the application of adversarial techniques also point to the broader issue of inconsistency and unpredictability in the current AI detection capabilities. The effectiveness of these techniques varies dramatically across detectors, suggesting that the internal algorithms and heuristics of these detectors are tuned differently and react distinctively to similar inputs. Therefore, the results even within an institution may differ depending on the tool being employed and how it is being used.

### Effectiveness of adversarial techniques

When exploring individual adversarial techniques, we see that relatively simple manipulations of content (such as the addition of spelling errors and increases in burstiness) are highly effective in evading detection, highlighting the limitations of AI detectors in distinguishing between human-like irregularities in text production and actual human writing. The least effective technique, 'increase complexity', had a minimal impact on detectability, which implies that dense and complex text alone is not a sufficient criterion for AI-generated content to pass undetected.

The results also indicated a significant variance in the detectability of content produced by different GenAI tools, with the output from Bard being easier to detect than Claude 2 and GPT-4. This suggests that not all GenAI outputs are equally recognisable by AI detectors, which has implications for their use in educational settings. The lower detectability of the content generated by Claude 2 and GPT-4 could make them more appealing to those who intend to circumvent academic honesty policies. Claude 2's superior performance in evading detection across all categories suggests that its outputs might align more closely with the nuanced and variable patterns of human writing or that it may be better at producing more human-like text in its original outputs.

**Implications of adversarial techniques for inclusive education in the age of AI**

The effectiveness of adversarial techniques in decreasing the likelihood of detecting machine-generated text raises concerns about both academic integrity and inclusivity. The current ability of GenAI tools to generate content that closely resembles human writing poses a significant threat to the fairness and authenticity of academic assessment. If students were to use AI tools to produce their academic work using these adversarial techniques, this would undermine the educational value and compromise the integrity of the assessment process, thereby benefiting some groups of students over others, especially those with access to paid GenAI tools and technical and procedural knowledge. Simultaneously, the fact that detectors may pose additional risks to NNESs that may produce text with less burstiness suggests that the risks of AI detection outweigh the potential benefits by exacerbating the existing inequities in academia. For instance, students with access to advanced AI tools and knowledge of adversarial techniques could gain an unfair advantage over their peers, further widening extant digital inequalities and the digital divide (Lutz, 2019).

The high level of variability between different detectors suggests that if institutions or academics take an individualistic approach to using these technologies (e.g. different academic staff at the same university using different GenAI detection technologies), the rates of efficacy would vary. Educators therefore need to be cognizant of the fact that while we focus on problematising text-detection software from a perspective of inclusivity, too much focus on GenAI as an emerging risk to academic integrity (Cotton et al., 2023; Perkins, 2023) may lead us to forget that that extant threats to inclusivity and educational equity, such as contract cheating or 'traditional' plagiarism, have not disappeared.

Error analysis provided insights into the risk of undetected cases and false accusations. With the rate of false accusations at 15%, considering the major impact that this could have on student outcomes, we consider this to be a major concern for student equity. Although some detectors did not have any false accusations, this appeared to come at the cost of a higher UDC ratio, indicating that many instances of AI-generated content could go unnoticed, potentially providing an unfair advantage to dishonest students who can apply these adversarial techniques in a matter of seconds to hide the true source of text.

**The use of GenAI text detectors in education**

While GenAI tools have great promise in enhancing academic practices for both students and teachers alike, and most publishing houses now permit the transparent use of such tools (Perkins & Roe, 2024a), there are circumstances in which the detection of GenAI texts remains necessary, such as in assessment practices which do not allow for the additional use of these tools or in computer-aided examinations. However, the fact that AI detection tools can be easily manipulated or bypassed using relatively simple adversarial techniques calls into question their viability as tools for maintaining assessment security.

Overall, our results demonstrate the challenges of current AI text detection tools being able to accurately determine whether a given piece of text was created by a human or a GenAI tool. This ability is further reduced when adversarial techniques are used to obscure the nature of a sample. If the goal of any given HEI was to use AI text detectors solely to determine whether a student has breached academic integrity guidelines, we would caution that the accuracy levels we have identified, coupled with the risks inherent in false accusations, means that we cannot recommend them for this purpose. This is not because of the demonstrated abilities of any one tool tested, as we recognise that developers are continuously updating these tools, and the detection of AI-generated content when subject to adversarial techniques is likely to improve. However, simultaneously, advances are being made in the development of more capable FMs that can produce more human-like content, resulting in a constant arms race between FMs and AI text detectors, with student inclusivity paying the price.

We believe that there will be a continued adjustment to GenAI-related learning and assessment practices within HE, with a move towards more acceptance of GenAI-enabled writing. Despite their shortcomings, AI text detectors may be part of a broader process of enabling discussions with students regarding the effective use of GenAI tools in academic writing. Students need experience and practice in learning how to work with GenAI tools, whether they are text-based or multimodal in nature. Being able to demonstrate to students how GenAI text may be integrated with their own writing is a task that these software tools may help with; however, this would only be possible if a non-punitive approach was taken to help students understand how GenAI tools can be used to support their own learning experiences.

**GenAI and the future of inclusive education**

Committing to inclusive methods of teaching, learning, and assessment implies a willingness to critique and adjust practices which incorporate new technologies. Similar to blended learning and communities of enquiry that contribute to the reorientation of the traditional classroom structure (Hilliard & Stewart, 2019), the presence of multiple forms of media on the Internet prompts us to reconsider the reliability and usefulness of source materials in the practice of academic argumentation (Radia & Stapleton, 2009). GenAI tools offer an opportunity to reconsider the traditional notions of misconduct and the potential barriers and inequities that a punitive approach to detection can face, particularly when relying on novel technologies.

New forms of academic fraud, such as online diploma mills, are growing (Roe & Perkins, 2023), and complex cases of fraud, such as contract cheating, continue to create unequal opportunities and outcomes (Curtis & Clare, 2024; Daly & Ryan, 2024); thus, efforts to promote inclusive forms of participation in higher education cannot solely focus on GenAI, as traditional, long-standing issues in equality of assessment have not gone anywhere, and some are even increasing.

Therefore, we need to foster alternative approaches to educational assessment which recognise the growing ubiquity of GenAI and account for obstacles to inclusive assessment designs. This requires fundamentally rethinking how we assess student learning, moving away from the traditional assessment methods that are easily compromised by AI tools. We would recommend that educators consider how GenAI tools can be more deeply integrated into learning and assessment so that students gain a deeper understanding of the practical and ethical applications of these technologies prior to moving into the workforce. Strategies such as the AI Assessment Scale (Perkins, Furze, et al., 2023), where the expected use of AI in any given assessment is clearly stated, could help guide educators in adapting assessment practices to align with the AI era, potentially leading to increased student outcomes and pedagogical redesign (Furze et al., 2024).

# Limitations

This study has several methodological limitations that should be considered when interpreting the findings and assessing their generalisability, which stem from both the choices made in the experimental design and the rapidly evolving nature of the AI landscape investigated.

First, the scope of our research was restricted to a relatively small number of samples, three GenAI tools, and seven AI detectors which do not encompass the entire spectrum of possible writing styles, types of manipulation, or available tools to generate and detect samples. The pace at which both GenAI tools and AI text detectors are developed and existing ones are updated means that our results are valid only for a specific snapshot in time.

Therefore, the findings here should be viewed as indicative rather than exhaustive, and subsequent research would benefit from a broader evaluation of the currently available detection software, as well as considering how some of the newest FMs, such as Claude 3 Opus and Gemini Pro 1.5, perform on these tasks.

Second, our experimental design did not mirror the iterative process that is likely to occur during the use of GenAI-supported academic writing and editing, particularly if there is a deliberate attempt to evade detection. In practice, students are likely to apply several of the techniques we have discussed and engage in continuous software-aided and manual refinement and adjustments, meaning that the ability of AI text-detection tools to detect this type of hybrid text would likely be even lower than demonstrated in the study. Future methodological designs should attempt to replicate the iterative nature of writing to provide more accurate assessments of AI detectors in academia as opposed to experimental settings.

Finally, some of the samples generated after applying adversarial techniques for testing may not accurately represent the quality of work that students would submit in a real-world setting. Although these samples evaded detection by software tools, they are likely to evoke suspicion from human markers because of their poor quality, strange phrasing choices, and excessive errors. This was particularly the case for some of the samples generated by either Claude 2 or Bard when asked to add spelling errors. Again, a broader sample of real-life cases would help explore this in more detail.

## Conclusion

The results of this study revealed that the average accuracy of AI text detectors in identifying non-manipulated AI-generated content was 39.5%, with a 67% accuracy rate for human-written control samples. When adversarial techniques were applied to the AI-generated samples, the average accuracy of the detectors dropped further to 22.14%, with some techniques, such as adding spelling errors and increasing burstiness, proving highly effective in evading detection. Error analysis also highlighted the risk of false accusations and undetected cases. These findings underscore the limitations of current AI text-detection tools in accurately determining the authorship of a given piece of text, particularly when faced with deliberate attempts to obscure the nature of the sample.

These findings demonstrate that the use of GenAI text-detection software has important ramifications for inclusivity, equality, and integrity in the AI era. Such tools are based on the assumption of assuring assessment fairness, which is a key principle in inclusive assessment design (Morris et al., 2019). However, our results showed that detectors are highly sensitive to the application of adversarial techniques, which may represent a more pragmatic view of how GenAI tools are used in academic writing. The implications of this are that, first, those with technological sensibility and resources, as well as inclination, will be able to disguise GenAI content relatively easily for the purpose of misrepresenting authorship. This reduces the equality of assessment (or publication of research) and thus, advantages some groups over others, acting as a barrier to inclusion. Furthermore, the potential for such tools to unfairly penalise students who write at a higher level of perplexity (including NNESs or lower-proficiency English speakers) suggests that, at the present time, the use of GenAI detection software for the identification of academic misconduct may produce barriers to inclusive assessment, rather than reduce them.

### Declaration of Generative AI and AI-assisted technologies in the writing process

During the preparation of this work, the authors used Claude 3 and GPT-4 to develop draft text, revise wording and adjust text throughout the manuscript. After using these tools, the authors reviewed and edited the content as needed and take full responsibility for the content of the publication.

**Acknowledgements**

The authors would like to acknowledge the contributions of Viet Anh Nguyet, who assisted with reviewing the literature, providing samples of human written text, and performing tests.

**Data availability statement:**

The data that support the findings of this study are openly available in Mendeley Data https://data.mendeley.com/datasets/xv6fk2mmh9/2 (Perkins et al., 2024)

**Declaration of interest:**

The authors have no relevant financial or non-financial interests to disclose.